\documentclass[aps,preprint,showkeys,amsmath,amssymb,footinbib,bibnotes]{revtex4-1}

\usepackage{bm}
\usepackage{graphicx,color}
\usepackage[T1]{fontenc}

\usepackage{bbold}
\usepackage{bbm}
\usepackage{array}
\usepackage{multirow}
\usepackage{bigstrut}
\usepackage{url}
\usepackage{braket}
\usepackage{color}

\def\mi{\mathrm{i}}
\def\cH{{\mathcal H}}

\def\bp{{\bf p}}
\def\br{{\bf r}}
\def\bk{{\bf k}}
\def\bsigma{{\boldsymbol\sigma}}
\def\bepsilon{{\boldsymbol\varepsilon}}

\begin{document}

\title{Theory of invariants-based formulation of $\bk\cdot\bp$ Hamiltonians with application to strained zinc-blende crystals}

\author{Johannes Wanner}
\author{Ulrich Eckern}
\email{ulrich.eckern@physik.uni-augsburg.de}
\author{Karl-Heinz H\"ock}
\affiliation{Universit\"at Augsburg, Institut f\"ur Physik, 86135 Augsburg, Germany}

\begin{abstract}

Group theoretical methods and $\bk\cdot\bp$ theory are combined to determine spin-dependent contributions to the effective conduction 
band Hamiltonian. To obtain the constants in the effective Hamiltonian, in general all invariants of the Hamiltonian have to be determined. 
Hence, we present a systematic approach to keep track of all possible invariants and apply it to the $\bk\cdot\bp$ Hamiltonian of crystals
with zinc-blende symmetry, in order to find all possible contributions to effective quantities such as effective mass, $g$-factor and 
Dresselhaus constant. Additional spin-dependent contributions to the effective Hamiltonian arise in the presence of strain. In particular, 
with regard to the constants $C_3$ and $D$ which describe spin-splitting linear in the components of $\bk$ and $\bepsilon$, considering 
all possible terms allowed by symmetry is crucial. 

\end{abstract}


\maketitle

\section{Introduction}
\label{sec_intro}

An effective description of bulk electrons and holes is crucial for the development and understanding of semiconductor based devices. 
The standard model for the calculation of the band structure $E(\bk)$ close to the fundamental band gap is the $\bk\cdot\bp$ 
theory \citep{Luttinger1955, Cardona1988, Novik2005, Kalinin2013, Wang1996, Nemanja2016}, or for quantum heterostructures its 
generalization, the envelope-function approximation. The full $\bk\cdot\bp$ Hamiltonian yields the exact description of the 
carriers in a periodic potential. Since it is an infinite-dimensional matrix, approximations are inevitable in order to diagonalize it. 
A very successful method is the limitation to bands close to the fundamental band gap. The most prominent ones are the $8\times 8$ 
and the $14\times 14$ Kane models, which include the valence and conduction band, and a second conduction band, respectively \citep{Kane1957}. 
Contributions from distant bands are usually taken into account by means of quasi-degenerate perturbation theory 
(``L\"owdin partitioning'' \citep{Loewdin1951}; a brief summary of this approach is given in 
appendix B of Ref.~\onlinecite{Winkler2003}) in second order.
For the description of larger regions in the first Brillouin zone more bands have to be taken into account, and higher-band models, 
like a 30-band model \citep{Cardona1966, Richard2004, Nemanja2016}, have been developed. 

The effect of strain on the band structure was first examined by Bir and Pikus \citep{Bir1974} within a 6-band model, and later 
generalized to higher-band models which also include strain-dependent terms originating from the spin-orbit interaction \citep{Bahder1990, Michelini2009}. 

In order to diagonalize such a finite-dimensional $\bk\cdot\bp$ Hamiltonian, all matrix elements have to be determined. Due to the 
crystal symmetry, not all of them are independent, and by applying group theoretical methods, identical as well as vanishing matrix 
elements can be identified. Within $\bk\cdot\bp$ theory, the non-vanishing materix elements (``invariants'') are considered as parameters; 
they have to be obtained experimentally or by different theories, e.g., density functional theory. Although the numerical diagonalization 
of finite $\bk\cdot\bp$ Hamiltonians with a number of bands larger than 30 is a trivial task for today's computers, increasing the number 
of bands leads to an increasing number of independent invariants. These invariants, however, are usually not easily obtained. Thus, 
instead of increasing the number of bands, remote bands are often taken into account by quasi-degenerate perturbation theory. 

In simple $\bk\cdot\bp$ models the spin-orbit interaction is neglected, and all invariants are given by the non-vanishing matrix 
elements of the momentum operator $\bp$. In more elaborated models including spin-orbit interaction, electromagnetic fields, or strain, 
a considerable number of additional invariants has to be considered. Hence, in section \ref{sec_inf_theory} we present a systematic 
approach which is independent of the selected symmetry group to keep track of all possibilities. Subsequently, in 
sections \ref{sec_eff_Ham_Td}--\ref{sec_s-cond_band} we apply this formulation to the $\bk\cdot\bp$ Hamiltonian of the zinc-blende 
structure to re-derive the effective mass $m_c$ and $g$-factor of the conduction band electrons, as well as the constants $C_3$ and $D$ 
describing strain-dependent spin-splitting. A brief conclusion is given in section \ref{sec_conclusion}. Some technical details are 
presented in three appendices.

The partitioning technique in quasi-degenerate perturbation theory was developed in a series of papers by L\"owdin, see 
Ref.~\onlinecite{Loewdin1968} and references therein. While our focus in this work is on his approach,
which is predominantly used in the context of $\bk\cdot\bp$ theory, we also wish to mention other relevant 
developments \cite{Brandow1967,Lindgren1974}. In brief, Brandow's \cite{Brandow1967} and Lindgrens's \cite{Lindgren1974} approaches
can be characterized as generalizations of Brillouin-Wigner and Rayleigh-Schr\"odinger perturbation theory, respectively. Differences
between the various schemes become apparent only in higher order. In particular, Lindgren's expansion obeys the linked-cluster
theorem in each order which implies the correct particle-number scaling in any finite order. A detailed discussion and comparison of
Brandow's and Lindgren's schemes, as well as an application to a relevant correlated-electron model, was presented recently 
\cite{Fresard2012}.

\section{Theory}
\label{sec_inf_theory}

In this work we concentrate on the influence of symmetry on various effective Hamiltonians within the framework of $\bk\cdot\bp$ theory. 
First, we briefly recall some group theoretical concepts, but refer the reader to the literature 
(Refs.~\onlinecite{Ludwig1996, Bradley2010, Dresselhaus2008, Altmann1994}) for more details.
As a start we consider the general Hamiltonian
\begin{equation}
\hat{\mathcal{H}} = \sum\limits_{o \mu\gamma } \mathcal{K}^{\mu}_{o; \gamma} \hat{\mathcal{O}}^{\mu}_{o; \gamma},
\label{eq_hamiltonian}
\end{equation}
where $\mu$, $\gamma$, and $o$ refer to the irreducible representation, the components of the operator with respect to the irreducible
represenation, and the repetition index: the latter is introduced to describe cases where the same irreducible representation appears
more than once.
The operators $\hat{\mathcal{O}}^{\mu}_{o; \gamma}$ transform according to
\begin{equation}
\hat{P}_g \hat{\mathcal{O}}^{\mu}_{o; \gamma}\hat{P}_g^{-1} = 
\sum\limits_{\gamma'}\mathcal{D}^{\Gamma^{\mu}}_{\gamma'\gamma} (g) \hat{\mathcal{O}}^{\mu}_{o; \gamma'}
\end{equation}
under the symmetry operation $\hat{P}_g$ of the group element $g$, and $\mathcal{D}^{\Gamma^{\mu}}_{\gamma'\gamma} (g)$ \footnote{The 
matrix representation $\mathcal{D}^{\Gamma^{\mu}}_{\gamma'\gamma} (g)$ can be found in textbooks such as Ref.~\onlinecite{Bradley2010}.} 
denotes the corresponding matrix representation of the irreducible representation $\Gamma^\mu$. 
The pre-factors $\mathcal{K}^{\mu}_{o; \gamma}$ 
-- depending on the Hamiltonian under consideration -- either are a constant or given by the external parameters such as the components of the wave 
vector $\bk$ or the strain tensor $\bepsilon$.
Moreover, we choose as a basis $\ket{\Gamma^{\alpha}_{i}(\delta)}$, where all degenerate states with band index $i$ transform under the 
symmetry operation $\hat{P}_g$ according the same irreducible representation $\Gamma^{\alpha}$:
\begin{equation}
\hat{P}_g \Ket{\Gamma^\alpha_i(\delta)} =\sum\limits_{\delta'}\mathcal{D}^{\Gamma^\alpha}_{\delta'\delta}(g)\Ket{\Gamma^\alpha_i(\delta')}. 
\label{eq_spatial_basis}
\end{equation}

Symmetry arguments imply that only those terms of the matrix elements 
$\cH^{\alpha\beta}_{ij;\delta\delta'}=\bra{\Gamma^{\alpha}_{i}(\delta)}\hat{\cH}\ket{\Gamma^{\beta}_{j}(\delta')}$ which transform according 
to $\Gamma^1 $ are non-zero. Furthermore, as the Hamiltonian has to remain invariant under all symmetry operations, not all matrix elements 
of $\hat{\cH}$ are independent. According to the generalized Wigner-Eckart theorem \citep{Ludwig1996}, each matrix element of the 
Hamiltonian $\hat{\cH}$ can therefore be decomposed as follows:
\begin{equation}
\cH^{\alpha\beta}_{ij;\delta\delta'}=\sum\limits_{o\mu \gamma}\mathcal{K}^{\mu }_{o; \gamma}\,\mathcal{O}^{\alpha\beta; \mu }_{o;\gamma;ij;\delta\delta'} = \sum\limits_{o\mu \gamma }\mathcal{I}^{\alpha\beta; \mu}_{o;ij}\,\mathcal{K}^{\mu}_{o; \gamma}\,\mathrm{X}^{\alpha\beta,\mu }_{\gamma;\delta\delta'},\label{eq_Ham0_elements}
\end{equation}
i.e., into a product of an invariant $\mathcal{I}^{\alpha\beta; \mu}_{o;ij}$, which is independent of $\gamma$, $\delta$, and $\delta'$,
and $\mathrm{X}^{\alpha\beta;\mu }_{\gamma;\delta\delta'}$ which denotes the Clebsch-Gordan coefficients. 
In the above, $(\alpha,i,\delta)$ and $(\beta,j,\delta')$ are the matrix indices, written in the appropriate compound form.
Generally a multiplicity index is required in Eq.~(\ref{eq_Ham0_elements}) (and has been taken into account in appendix \ref{App_Def_X}
where further details about the matrices X are given). In order to keep the notation ``lean'', we have dropped this index in the main
text, as it would appear only in the explicit expressions given at the end of section \ref{sec_eff_Ham_Td}, namely 
Eqs.~(\ref{cubicDresselhaus}) and (\ref{EffHam_Dress_eq_lambda}).
The invariants can be obtained by
\begin{equation}
\mathcal{I}^{\alpha\beta; \mu}_{o;ij}=\sum\limits_{\delta\delta'}\mathcal{O}^{\alpha\beta; \mu }_{o;\gamma;ij;\delta\delta'}\,\mathrm{X}^{\beta\alpha;\mu }_{\gamma;\delta'\delta}. \label{eq_I_inv}
\end{equation}
For an infinite number of bands the matrix $\mathcal{I}^{\alpha\beta; \mu}_{o}$ is infinite-dimensional, and contains all relevant 
information of the operators $\hat{\mathcal{O}}^\mu_{o;\gamma}$. On the other hand, the matrices $\mathrm{X}^{\alpha\beta;\mu }_{\gamma}$ 
depend only on the selected symmetry class. Therefore the expression (\ref{eq_Ham0_elements}) represents a convenient separation of the
material-dependent quantities $\mathcal{I}^{\alpha\beta; \mu}_{o}$ from the parameter-dependent matrices $\mathcal{K}^{\mu}_{o;\gamma}\,\mathrm{X}^{\alpha\beta;\mu }_{\gamma}$, which contain the information on the symmetry group. This separation can be directly translated into each block of the Hamiltonian, 
\begin{equation}
{\cH}^{\alpha\beta}=\sum\limits_{ o\mu \gamma}\mathcal{I}^{\alpha\beta;\mu}_{o}\otimes\left(\mathcal{K}^{\mu}_{o;\gamma}\,\mathrm{X}^{\alpha\beta;\mu}_{\gamma}\right),\label{eq_hamab}
\end{equation}
where $\otimes$ denotes the direct product. Note that Eq.~(\ref{eq_hamab}) is a short-hand version of Eq.~(\ref{eq_Ham0_elements}).
The fact that this separation keeps its form even after applying quasi-degenerate L\"owdin perturbation theory \citep{Loewdin1951,Loewdin1968}, 
renders it especially convenient for calculating, e.g., the effective mass or the effective $g$-factor of the charge carriers. Up to 
second-order, the Hamiltonian for the degenerate $\Gamma^{\alpha}_{i}$ bands is then given by the following expression:
\begin{align}
\hat{H}^{\alpha}_{i}= &E_{\alpha; i}+ \sum\limits_{o\mu\gamma}\mathcal{I}^{\alpha\alpha; \mu}_{o;ii}\,\mathcal{K}^{\mu}_{o; \gamma}\,\mathrm{X}^{\alpha\alpha;\mu }_{\gamma}+\nonumber \\
& + \sum\limits_{\substack{o\mu \gamma  \\o' \mu' \gamma'  \\ \beta\, j\neq i}}\dfrac{\mathcal{I}^{\alpha\beta ;\mu}_{o;ij}\mathcal{I}^{\beta\alpha;\mu'}_{o';ji}}{E_{\alpha;i}-E_{\beta;j}}\,\left(\mathcal{K}^{\mu }_{o;\gamma}\mathcal{K}^{\mu' }_{o';\gamma'}\,\mathrm{X}^{\alpha\beta;\mu }_{\gamma}\mathrm{X}^{\beta\alpha;\mu'}_{\gamma'}\right) +\ldots \label{eq_Ham_eff}
\end{align}

\section{Effective conduction band Hamiltonians without strain}
\label{sec_eff_Ham_Td}

Each symmetry of a crystal structure corresponds to a specific feature in the energy spectrum, i.e., at points of symmetry bands are 
degenerate. In semiconductors, the splitting of energy bands in the vicinity of such points can efficiently be described within $\bk\cdot\bp$ 
theory which involves an expansion of the energy spectrum in terms of the Bloch states of this symmetry point. Instead of calculating the 
energy spectrum $E_i(\bk)$ of each band independently, within $\bk\cdot\bp$ theory a different approach commonly is used. With the help 
of quasi-degenerate perturbation theory, 
Eq.~(\ref{eq_Ham_eff}), all bands of the considered energy spectrum are decoupled from remote bands resulting in effective 
Hamiltonians $\hat{H}_i(\bk) $. The energy spectrum can then be obtained by diagonalizing $\hat{H}_i(\bk) $. 

We apply the general considerations of section \ref{sec_inf_theory} to the $\bk\cdot\bp$ Hamiltonian of an electron in a crystal with 
zinc-blende structure. Information on the corresponding symmetry group $T_d \otimes \mathcal{D}^{1/2}$ can be found in the literature 
(see, e.g., Refs.~\onlinecite{Wanner2016,Bradley2010,Koster1963,Altmann1994,Trebin1979,Winkler2003}).

Without strain, the Hamiltonian is given by \cite{Dresselhaus1955,Kane1957,Suzuki1974,Bahder1990,Winkler2003}:
\begin{equation}
\hat{\cH}=\hat{\cH}_{0}+\hat{\cH}_{\mathrm{so}}+\hat{\cH}_{\bk}+\hat{\cH}_{\bk\cdot\bp}+\hat{\cH}'_{\mathrm{so}}, \label{eq_KPHam}
\end{equation}
with
\begin{align}
\hat{\cH}_{0}&=\dfrac{\bp^2}{2m_0}+V_0(\mathbf{x}),  \\
\hat{\cH}_{\mathrm{so}}\equiv \hat{\cH}_{\mathrm{I}}&= \dfrac{\hbar}{4 m_0^2 c^2}(\nabla V_0 )\times\bp\cdot \bsigma,\\
\hat{\cH}_{\bk}\equiv \hat{\cH}_{\mathrm{II}}&=\dfrac{\hbar^2 \bk^2}{2m_0},\\
\hat{\cH}_{\bk\cdot\bp}\equiv \hat{\cH}_{\mathrm{III}}&=\dfrac{\hbar}{m_0}\bp\cdot \bk ,\\
\hat{\cH}'_{\mathrm{so}}\equiv \hat{\cH}_{\mathrm{IV}}&=\dfrac{\hbar^2}{4 m_0^2 c^2}(\nabla V_0 )\times\bk\cdot \bsigma. 
\end{align}

In order to apply perturbation theory, one has to split the above Hamiltonian in an unperturbed part and a small perturbation. In the literature
\cite{Kane1957,Bahder1990,Winkler2003}
this is often done by declaring $\hat{\cH}_{0}$ the unperturbed part and the rest as perturbation. Unfortunately, this choice has the 
disadvantage that it leads to higher-order contributions to effective quantities such as the effective mass, due to a coupling between $\hat{\cH}_{0}$ 
and the spin-orbit part $\hat{\cH}_{\mathrm{so}}$. To avoid these complications, we choose $\hat{\cH}_{0} + \hat{\cH}_{\mathrm{so}} $ as 
unperturbed part, and
\begin{equation}
\left(\hat{\cH}_{0} + \hat{\cH}_{\mathrm{so}} \right)\Ket{\Gamma^{\alpha}_{i}(\delta)} = E_{\alpha;j}\Ket{\Gamma^{\alpha}_{i}(\delta)}.
\end{equation}

Let us now consider the effective Hamiltonian of the bands with $\Gamma^6$ symmetry, with the most prominent representative being the 
conduction band. The trivial Hamiltonian $\hat{\cH}_{\mathrm{II}}$ is already diagonal, hence we need only pay attention 
to $ \hat{\cH}_{\mathrm{III}}$ and $ \hat{\cH}_{\mathrm{IV}}$. Both Hamiltonians contain the same 
parameter $\mathcal{K}^{5}_{\mathrm{III}}=\mathcal{K}^{5}_{\mathrm{IV}} = \bk $, thus we combine their operators as follows:
\begin{align}
\hat{\mathcal{O}}^{5}_{\bk}\equiv \hat{\mathcal{O}}^{5}_{\mathrm{III}} + \hat{\mathcal{O}}^{5}_{\mathrm{IV}} = \dfrac{\hbar}{m_0}\bp +\dfrac{\hbar^2}{4 m_0^2 c^2}\bsigma\times(\nabla V_0 ),
\end{align}
which transform according to $\Gamma^{5}$ . 

For the conduction band in first-order perturbation theory, symmetry allows only operators which transform according to $\Gamma^1$ 
or $\Gamma^4$. Hence, the first contributions from $ \hat{\cH}_{\mathrm{III}}$ and $ \hat{\cH}_{\mathrm{IV}}$ to the effective 
Hamiltonian of a $\Ket{\Gamma^6_i}$ band arise only in second-order perturbation theory, and are given by ($\mathbf{B}=\mi\hbar \bk\times\bk/e$):
\begin{equation}
\hat{H}^6_i = E_{6;i} +  \dfrac{\hbar^2\bk^2}{2m_i} + \dfrac{g_i}{2}\mu_{\mathrm{B}}\mathbf{B}\cdot\bsigma,
\end{equation}
where
\begin{align}
\dfrac{\hbar^2 }{2m_{i}} &=\dfrac{\hbar^2 }{2m_{0}} +\frac{1}{2}\sum\limits_{j\neq i} \left( \dfrac{\mathcal{I}^{67; 5}_{\bk;ij}\mathcal{I}^{76; 5}_{\bk; ji}}{E_{6;i}-E_{7;j}}+\dfrac{\mathcal{I}^{68; 5 }_{\bk;ij}\mathcal{I}^{86; 5}_{\bk;ji}}{E_{6;i}-E_{8;j}}\right), \label{Gamma6_eq_eff_m}
\end{align}
which defines the effective mass, and
\begin{align}
g_i =\frac{2 m_0}{\hbar^2} \sum\limits_{j \neq i} \left( \dfrac{\mathcal{I}^{67; 5}_{\bk;ij}\mathcal{I}^{76; 5}_{\bk; ji}}{E_{6;i}-E_{7;j}}-\dfrac{1}{2}\dfrac{\mathcal{I}^{68; 5}_{\bk;ij}\mathcal{I}^{86; 5}_{\bk;ji}}{E_{6;i}-E_{8;j}}\right),
\end{align}
the effective $g$-factor. Hence, up to second order -- without an external magnetic field -- there is no spin-splitting for bands with $\Gamma^6$ 
symmetry. However, a spin-dependent splitting arises in higher order. The lowest-order term that leads to such a spin-splitting is called 
the cubic Dresselhaus term \citep{Dresselhaus1955}:
\begin{align}
\hat{H}^{\mathrm{D}}_c &=\lambda_{i}[k_x \left(k_y^2-k_z^2 \right)\sigma^x\nonumber \\
					&\phantom{=}+k_y \left(k_z^2-k_x^2 \right)\sigma^y +k_z \left(k_x^2-k_y^2 \right)\sigma^z ] .
\end{align}
This contribution to the effective Hamiltonian couples the motion of the electrons in $\Gamma^6$ bands to their spin. The structure 
of Eq.~(\ref{eq_hamab}) is very convenient for finding the relevant matrix products which are responsible for the cubic Dresselhaus 
term. These are:
\begin{align}
k_i k_j k_l \mathrm{X}^{67;5}_{i}\mathrm{X}^{78; 5}_{j}\mathrm{X}^{86; 5}_l , \\
k_i k_j k_l \mathrm{X}^{68;5}_i \mathrm{X}^{87;5}_j \mathrm{X}^{76;5}_l , \\
k_i k_j k_l \mathrm{X}^{68;5}_i \mathrm{X}^{88;5(b)}_{j}\mathrm{X}^{86;5}_l , \label{cubicDresselhaus}
\end{align}
where Einstein's summation convention is implied. Note that for the derivation of these matrix products, no information about the 
invariants is necessary. Moreover, this selection of the relevant matrix products is directly conferred upon the invariants. The 
material constant of the Dresselhaus term is thus readily given by 
\begin{align}
\lambda_{i}=& \frac{1}{2}\sqrt{\frac{3}{2}}\sum\limits_{\substack{ jll'}}\left[\dfrac{\mi\left(\mathcal{I}^{67; 5}_{\bk; ij}\mathcal{I}^{78 ; 5}_{\bk; jl}\mathcal{I}^{86 ; 5}_{\bk; li}-\mathcal{I}^{68; 5}_{\bk; il}\mathcal{I}^{87 ; 5}_{\bk; lj}\mathcal{I}^{76 ; 5}_{\bk; ji}\right)}{\left(E_{6, i}-E_{7, j}\right)\left(E_{6, i}-E_{8, l}\right)} \right.\nonumber \\
&+\left.\frac{1}{2\sqrt{2}}\dfrac{\mathcal{I}^{68; 5}_{\bk; il}\mathcal{I}^{88; 5(b)}_{\bk; ll'}\mathcal{I}^{86 ; 5}_{\bk; l'i}}{\left(E_{6, i}-E_{8, l}\right)\left(E_{6, i}-E_{8, l'}\right)} \right],\label{EffHam_Dress_eq_lambda}
\end{align}
where once again the expression holds for arbitrary bands with $\Gamma^6 $ symmetry. See appendix \ref{App_Def_X} for an explanation of the superscript
``(b)'' in Eqs.~(\ref{cubicDresselhaus}) and (\ref{EffHam_Dress_eq_lambda}).

\section{Strain induced spin-orbit splitting}
\label{sec_strain_spin_split}

In this section we include strain effects which lead, in addition to the Dresselhaus term, to spin-orbit coupling. In the presence 
of linear strain the symmetry of the crystal and thus of the potential $V_0(\br)\rightarrow V_{\bepsilon}(\br)$ is reduced. The 
symmetry of the unstrained crystal can, however, be restored by following the method of Bir and Pikus \citep{Bir1974, Bahder1990, Zhang1994} 
who instead of deforming the crystal consider a deformed coordinate system. The potential $V_{\bepsilon}$ then has the same periodicity 
as the potential in the absence of strain, and can be expanded in terms of $\bepsilon$. As a consequence of the coordinate transformation, 
there are additional strain-dependent terms in the Hamiltonian, $\hat{H}\rightarrow\hat{H}+\hat{D}_0+\hat{D}_{\bk\cdot\bp}+\hat{D}_{\mathrm{so}}+\hat{D}'_{\mathrm{so}}$. Up to first order in strain they are \citep{Bahder1990}:
\begin{align}
\hat{D}_0 \equiv \hat{H}_{\mathrm{V}} &= \sum\limits_{ij} \left[-\frac{1}{m_0 }p_i p_j + V_{ij}(\br) \right]\varepsilon_{ij}\equiv \sum\limits_{ij} \hat{D}_{ij}\varepsilon_{ij}, \label{EffHam_STRAIN_eq_D0}\\
\hat{D}_{\mathrm{so}}\equiv\hat{H}_{\mathrm{VI}} &=\frac{\hbar}{4 m_0^2 c^2 } \left[\sum\limits_{ij}\varepsilon_{ij}\nabla V_{ij}(\br) \times \bp\cdot\bsigma \right. \nonumber \\
                        &\left. \phantom{=\frac{\hbar}{4 m_0^2 c^2 }=} -\left(\nabla V_{0}(\br)\cdot\bepsilon \right)\times\bp\cdot\bsigma\right. \label{ST_eq_Dso}\nonumber \\
                        &\left.\phantom{=\frac{\hbar}{4 m_0^2 c^2 }=}-\nabla V_{0}(\br)\times\left(\bepsilon\cdot\bp \right)\cdot\bsigma \vphantom{\sum\limits_{ij}}\right], 
\end{align}
\begin{align}                       
\hat{D}_{\bk\cdot\bp}\equiv \hat{H}_{\mathrm{VII}} &= -\frac{\hbar}{m_0 } \bp\cdot\bepsilon\cdot\bk,  \label{ST_eq_Dkp} \\                        
\hat{D}'_{\mathrm{so}}\equiv\hat{H}_{\mathrm{VIII}} &=\frac{\hbar}{4 m_0^2 c^2 } \left[\sum\limits_{ij}\varepsilon_{ij}\nabla V_{ij}(\br) \times \bk\cdot\bsigma\right. \nonumber \\
						&\left.\phantom{=\frac{\hbar}{4 m_0^2 c^2 }=} -\left(\nabla V_{0}(\br)\cdot\bepsilon \right)\times\bk\cdot\bsigma  \vphantom{\sum\limits_{ij}}\right], \label{EffHam_STRAIN_eq_Dso_prime}
\end{align}
with 
\begin{equation}
V_{ij}(\br)=\frac{1}{2-\delta_{ij}}\lim\limits_{\bepsilon\rightarrow 0} \frac{V_{\bepsilon}\left[(1+\bepsilon)\br\right]-V_{0}(\br)}{\varepsilon_{ij}}.
\end{equation}
Since all symmetry operations of the unstrained Hamiltonian still apply, the influence of strain on the energy spectrum $E(\bk, \bepsilon)$ 
can again be derived according to Eq.~(\ref{eq_Ham_eff}). Nevertheless, since the additional terms in the Hamiltonian are limited to first 
order in strain, we also restrict ourselves to terms linear in $\bepsilon$ in the effective Hamiltonian. In order to derive this effective 
Hamiltonian, additional invariants, and thus the symmetry properties of the operators of the strain-dependent Hamiltonians, have to be determined. 

Considering the symmetry properties of the strain tensor, there are three different symmetry adapted strain components:
\begin{align}
\mathcal{K}^{1}_{\bepsilon}	&= \mathrm{Tr}\,\bepsilon ,  \\
\mathcal{K}^{3}_{\bepsilon}	&= \left(\frac{2\varepsilon_{zz}-\varepsilon_{xx}-\varepsilon_{yy}}{\sqrt{6}},\frac{\varepsilon_{xx}-\varepsilon_{yy}}{\sqrt{2}} \right) ,  \\
\mathcal{K}^{5}_{\bepsilon}	&= (\varepsilon_{yz},	\varepsilon_{zx} ,\varepsilon_{xy}) .
\end{align}
In first order only operators with $\Gamma^1$ or $\Gamma^4$ symmetry need to be considered. As the strain tensor is symmetric, there is no operator $\hat{\mathcal{O}}^{4}_{\bepsilon}$ with $\Gamma^4$ symmetry, and only the operator $\hat{\mathcal{O}}^{1}_{\bepsilon}$ remains.
Hence, up to first order a hydrostatic strain-dependent contribution to the effective Hamiltonian
\begin{equation}
\hat{H}^{6}_{\bepsilon;i } = a_i \mathrm{Tr}\,\bepsilon
\end{equation}
with 
\begin{equation}
a_i = \mathcal{I}^{66, 1}_{\bepsilon,ii} 
\end{equation}
arises. As a consequence, linear strain is not able to lift the degeneracy of $\Gamma^6$ bands at the gamma point. 

Next we consider linear combinations of the $\bk$ vector and the strain tensor $\bepsilon$. With the three components of $\bk$, and the 
six independent ones of $\bepsilon$, there are a total of 18 different combinations $k_i \varepsilon_{jl}$. Thus there are also 18 independent 
components of the parameters $\mathcal{K}_{\bk\bepsilon;\gamma}^{\mu}$. Regarding the direct product of the irreducible representations 
$\Gamma^5 \otimes (\Gamma^1 \oplus \Gamma^3 \oplus \Gamma^5) = \Gamma^1 \oplus \Gamma^3 \oplus 2\Gamma^4 \oplus 3\Gamma^5$, there is 
either one combination transforming according to $\Gamma^1$ and $\Gamma^3$, two according to $\Gamma^4$, and three according to $\Gamma^5$. 
The parameters needed for the conduction band are:
\begin{align}
\mathcal{K}^{1}_{\bk\bepsilon} &=  \mathcal{K}^{5}_{\bk;\gamma} \mathrm{X}^{55; 1}_{\gamma\gamma '} \mathcal{K}^{5}_{\bepsilon;\gamma '}=\frac{\varepsilon_{yz}k_x +\varepsilon_{zx}k_y +\varepsilon_{xy}k_z}{\sqrt{3}} , \label{eqKe1} \\
\mathcal{K}^{4}_{\bk\bepsilon 1;\lambda} &=\mathcal{K}^{5}_{\bk;\gamma} \mathrm{X}^{53; 4}_{\lambda;\gamma\gamma '} \mathcal{K}^{3}_{\bepsilon;\gamma '} = \dfrac{1}{\sqrt{2}}\left( \begin{array}{c}
(\varepsilon_{yy}-\varepsilon_{zz} )k_x\\
(\varepsilon_{zz}-\varepsilon_{xx} )k_y \\
(\varepsilon_{xx}-\varepsilon_{yy} )k_z
\end{array}  \right), \label{eqKe41}\\
\mathcal{K}^{4}_{\bk\bepsilon 2;\lambda} &=\mathcal{K}^{5}_{\bk;\gamma} \mathrm{X}^{55; 4}_{\lambda;\gamma\gamma '} \mathcal{K}^{5}_{\bepsilon;\gamma '} = \dfrac{1}{\sqrt{2}} \bk\times\left(\begin{array}{c}
\varepsilon_{yz}\\
\varepsilon_{zx}\\
\varepsilon_{xy}
\end{array}   \right)\label{eqKe42} .
\end{align}
The resulting effective Hamiltonians therefore are:
\begin{equation}
\hat{H}^{6;1}_{\bk\bepsilon;i} =  b_{i} (\varepsilon_{yz}k_x +\varepsilon_{zx}k_y +\varepsilon_{xy}k_z),\label{eq_HamKe1}
\end{equation}
\begin{align}
\hat{H}^{6;4}_{\bk\bepsilon 1; i} = &D_{i} \left[(\varepsilon_{zz}-\varepsilon_{yy} )k_x \sigma^x\right.\label{eq_HamD} \nonumber \\
&\left.+(\varepsilon_{xx}-\varepsilon_{zz} )k_y\sigma^y +(\varepsilon_{yy}-\varepsilon_{xx} )k_z\sigma^z \right],
\end{align}
and
\begin{align}
\hat{H}^{6;4}_{\bk\bepsilon 2; i} =&\frac{1}{2} C_{3;i} \left[(\varepsilon_{xy} k_y-\varepsilon_{zx} k_z )\sigma^x\right. \nonumber \\
&\left.+(\varepsilon_{yz} k_z-\varepsilon_{xy} k_x )\sigma^y+(\varepsilon_{zx} k_x-\varepsilon_{yz} k_y )\sigma^z \right],\label{eq_HamC}
\end{align}
where we introduced the material constants $C_{3;i}$ and $D_{i}$ according to Ref.~\onlinecite{Bernevig2005}.

There are two possible ways for generating this type of effective Hamiltonians. One possibility is first-order perturbation theory with 
respect to the Hamiltonians $\hat{H}_\mathrm{VII}$ and $\hat{H}_\mathrm{VIII}$. The second source of such terms originates from 
second-order perturbation theory including combinations of the Hamiltonians $\hat{H}_\mathrm{III}$ and $\hat{H}_\mathrm{IV}$ 
with $\hat{H}_\mathrm{V}$ and $\hat{H}_\mathrm{VI}$, respectively. Starting with the first-order contributions originating 
from $\hat{H}_\mathrm{VIII}$, their corresponding operators are $\hat{\mathcal{O}}^{1}_{\bk\bepsilon}$, $\hat{\mathcal{O}}^{4}_{\bk\bepsilon 1;\lambda}$, 
and $\hat{\mathcal{O}}^{4}_{\bk\bepsilon 2;\lambda}$. Again, in first order the contributions from $\hat{H}_\mathrm{VII}$ vanish due to symmetry. 
With these operators given, the required invariants $\mathcal{I}^{66; 1}_{\bk\bepsilon ;ii}$, $\mathcal{I}^{66; 4}_{\bk\bepsilon 1;ii}$, 
and $\mathcal{I}^{66; 4}_{\bk\bepsilon 2;ii}$ follow directly from Eq.~(\ref{eq_I_inv}). 

According to Eqs.~(\ref{eqKe1})--(\ref{eqKe42}) the second-order contributions to the effective Hamiltonians 
(\ref{eq_HamKe1})--(\ref{eq_HamC}) are generated by a coupling of the invariants of $\hat{\mathcal{O}}^{5}_{\bk} $ with those 
of $\hat{\mathcal{O}}^{3}_{\bepsilon} $ and $\hat{\mathcal{O}}^{5}_{\bepsilon} $, respectively. For an arbitrary band 
with $\Gamma^6$ symmetry, the material constants are finally given by:
\begin{align}
b_{i} &=\frac{\mathcal{I}^{66; 1}_{\bk\bepsilon ;ii}}{\sqrt{6}} + \frac{1}{2}\sum\limits_{j\neq i} \left( \dfrac{\mathcal{I}^{67; 5}_{\bk;ij}\mathcal{I}^{76; 5}_{\bepsilon; ji}}{E_{6;i}-E_{7;j}}+\dfrac{\mathcal{I}^{68; 5 }_{\bk;ij}\mathcal{I}^{86; 5}_{\bepsilon;ji}}{E_{6;i}-E_{8;j}}\right),\label{eq_b_general}
\end{align}
\begin{align}
D_{i} &=-\frac{\mathcal{I}^{66; 4}_{\bk\bepsilon 1;ii}}{2}+ \frac{1}{2\sqrt{2}}\sum\limits_{j}\dfrac{\mathcal{I}^{68; 5}_{\bk;ij}\mathcal{I}^{86; 3 }_{\bepsilon; ji}+\mathcal{I}^{68; 3 }_{\bepsilon; ij}\mathcal{I}^{86; 5}_{\bk; ji}}{E_{6;i}-E_{8;j}},
\label{eq_D_general}
\end{align}
and
\begin{align}
C_{3; i} &=\mathcal{I}^{66; 4}_{\bk\bepsilon 2;ii}+ \mi\sum\limits_{j}\left(\dfrac{\mathcal{I}^{67; 5}_{\bk;ij}\mathcal{I}^{76; 5 }_{\bepsilon; ji}-\mathcal{I}^{67;5}_{\bepsilon;ij}\mathcal{I}^{76;5}_{\bk; ji}}{E_{6;i}-E_{7;j}}\right. \nonumber \\
&\left.\phantom{= \mi\sum}-\frac{1}{2}\dfrac{\mathcal{I}^{68;5 }_{\bk;ij}\mathcal{I}^{86;5}_{\bepsilon; ji}-\mathcal{I}^{68;5}_{\bepsilon; ij}\mathcal{I}^{86;5 }_{\bk;ji}}{E_{6;i}-E_{8;j}}\right).
\label{eq_C_general}
\end{align}
So far we considered general bands with $\Gamma^6 $ symmetry. If, in addition, time reversal invariance is imposed, see the next section,
it is apparent that $b_{i}$ vanishes; compare Eq.~(\ref{eq_HamKe1}).

\section{S-like conduction band}
\label{sec_s-cond_band}

The bands with $\Gamma^6$ symmetry considered so far (cf.\ appendix \ref{App_Basis}) consist in general of a spatial part with 
$\Gamma^1$ and $\Gamma^4$ symmetry. The conduction band, however, is usually assumed to be ``s''-like, and consists therefore 
only of a spatial part with $\ket{\Gamma^1_{c}}$ symmetry ($\eta_c^4 = 0$). Taking this restriction into account, not every 
term in the above operators is able to contribute to its corresponding invariant. Instead of discussing this point for every 
material constant derived above, we focus on the constants $D_{c}$ and $C_{3;c}$ which describe the coupling strength of a 
strain-dependent spin-splitting. 

To determine the non-vanishing terms, the operators $\hat{\mathcal{O}}^{\mu}_{o; \gamma}$ have to be separated into a ``spatial'' 
operator $\hat{\mathrm{O}}^{\mu}_{o; \gamma}$ and a Pauli matrix. The operators 
$\hat{\mathrm{O}}^{5}_{\mathrm{V};\lambda}\equiv\hat{\mathcal{O}}^{5}_{\mathrm{V};\lambda}$, $\hat{\mathrm{O}}^{3}_{\mathrm{V};\lambda}\equiv\hat{\mathcal{O}}^{3}_{\mathrm{V};\lambda}$, and $\hat{\mathrm{O}}^{5}_{\mathrm{V};\lambda}\equiv\hat{\mathcal{O}}^{5}_{\mathrm{V};\lambda}$ 
are already spin-independent, and the remaining operators can be found by decomposing them with the help of
\begin{equation}
\hat{\mathcal{O}}^{\mu}_{o; \gamma}  = \sum\limits_{\kappa\lambda s} \hat{\mathrm{O}}^{\kappa}_{o; \gamma}\sigma^{s}\mathrm{X}^{\kappa 4; \mu}_{\gamma;\lambda s} \label{EffHam_SymmHam_eq_SOI_O}
\end{equation}
into a product of a spatial operator $ \hat{\mathrm{O}}^{\kappa}_{o; \gamma}$ and a Pauli matrix.

First-order contributions to the conduction band are then only possible if an operator $\hat{\mathrm{O}}^{1}_{o; \gamma}$
which transforms according to $\Gamma^1$ is contained in the above decomposition. This is not the case for the operator 
$\hat{\mathcal{O}}^{4}_{\bk\bepsilon 1;\lambda}$, and the invariant corresponding to the conduction band constant $D_c$ 
vanishes, $\mathcal{I}^{66; 4}_{\bk\bepsilon 1;cc} = 0$. 
The invariant $\mathcal{I}^{66; 4}_{\bk\bepsilon 2;cc} $ of the second 
material constant $C_{3;c}$, however, is non-zero, and the corresponding spatial operator is given by
\begin{equation}
\hat{\mathrm{O}}^{1}_{\mathrm{VIII}} = -\frac{\sqrt{2}\hbar}{6 m_0^2 c^2}(\partial_x V_{yz}+ \partial_y V_{zx}+ \partial_z V_{xy}).
\end{equation}
As the first-order contribution is prohibited by symmetry, the material constant $D_c$ has to arise in second-order perturbation theory 
which couples the invariants of $\hat{\mathcal{O}}^{5}_{\bk;\lambda}$ to those of $\hat{\mathcal{O}}^{3}_{\bepsilon;\lambda}$. 

The operator $\hat{\mathcal{O}}^{5}_{\mathrm{IV}}$ contains only the spatial operator
\begin{equation}
\hat{\mathrm{O}}^{5}_{\mathrm{IV};\lambda}= \frac{\sqrt{2}\hbar^2}{4 m_0^2 c^2}\nabla V_0 
\end{equation}
in its decomposition. The spatial operators of $\hat{\mathcal{O}}^{5}_{\bk}$ thus both transform according to $\Gamma^5$, and 
the s-like conduction band couples only to bands with spatial symmetry $\Gamma^5$. According to Eqs.~(\ref{eq_D_general}) 
and (\ref{eq_C_general}) there are therefore only spatial operators with $\Gamma^5$ symmetry allowed for the invariants of 
$\hat{\mathcal{O}}^{3}_{\bepsilon;\lambda}$ and $\hat{\mathcal{O}}^{5}_{\bepsilon;\lambda}$. Both of them possess such operators:
\begin{align}
\hat{\mathrm{O}}^{5}_{\mathrm{VIa};1} = \frac{\sqrt{3}\hbar}{8m_0^2 c^2}\nabla(V_{yy}-V_{zz})\times \bp\cdot \hat{e}_x  \\
\end{align}
is contained in the decomposition of $ \hat{\mathcal{O}}^{3}_{\mathrm{VI};\lambda}$, and
\begin{equation}
\hat{\mathrm{O}}^{5}_{\mathrm{VIb};1} = \frac{\sqrt{2}\hbar}{4m_0^2 c^2}(\nabla V_{zx}\times \bp\cdot \hat{e}_z -\nabla V_{xy}\times \bp\cdot \hat{e}_y )  
\end{equation}
in $\hat{\mathcal{O}}^{5}_{\mathrm{VI};\lambda}$.

\section{Conclusion}
\label{sec_conclusion}

The general approach described in section \ref{sec_inf_theory} enables a systematic derivation of all invariants. This is 
particularly important for all finite-band models as well as for $\bk\cdot\bp$ models which include remote bands by 
quasi-degenerate perturbation theory. We demonstrated this by discussing the strain-dependent spin-splitting contributions 
to the effective conduction band Hamiltonian. 

As an important result, we find, in particular, that the constant $D_c$ can only arise from the invariants of 
$\hat{\mathrm{O}}^{5}_{\mathrm{VIa};1}$, i.e., a part of the first term of the Hamiltonian $\hat{H}_\mathrm{VI}$; this term, 
however, has been neglected in previous works \citep{Bahder1990, Zhang1994, Michelini2009}. Hence, in these previous studies 
there is no spin-splitting of the type (\ref{eq_D_general}), and $D_c =0$. In contrast, the experiments of 
Refs.~\onlinecite{Bernevig2005,Norman2010,Kato2004,Hruska2006} show that $D_c$ is non-zero, though by a factor of $\sim$\,100 
smaller than $C_3$. The experimental result thus indicates that either the invariants of $\hat{\mathrm{O}}^{5}_{\mathrm{VIa};1}$ 
are not negligible; or that the general assumption of a pure s-like conduction band is not valid. This question can, in principle, 
be decided theoretically by a more detailed calculation which would require, however, a precise determination of the single-particle 
pseudo potential $V_0(\br)$, e.g., based on density functional theory. The results of such a study, which is beyond the scope of 
the present work, would thus allow a direct calculation of the invariants based on the solution of the Hamiltonian  
$\hat{\cH}_{0} + \hat{\cH}_{\mathrm{so}}$.

In summary, this work shows that band parameters can be obtained in a convenient way by a consequent treatment in terms of 
symmetry arguments. This becomes especially obvious in the presence of strain where the invariants of higher-order tensor 
operators (in $ \hat{\cH}_{\mathrm{VI}}$, e.g., fourth order) have to be calculated.

\acknowledgments

Supported by the Deutsche Forschungsgemeinschaft through TRR 80.

\appendix

\section{Basis states}
\label{App_Basis}

In crystals with zinc-blende structure the eigenstates have to transform either according to $\Gamma^6, \Gamma^7$, or $\Gamma^8$, respectively. 
The states used in this work are given as follows:
\begin{align}
\Ket{\Gamma^{6}_{i}(1)} = &\eta^1_i  \Ket{\begin{array}{c}\Gamma^1_i \\ 0 \end{array}}-\frac{\eta^{4}_{i}}{\sqrt{3}}\Ket{\begin{array}{c} \Gamma^4_i(3) \\ \Gamma^4_i(1)+\mi \Gamma^4_i(2) \end{array}},  \\
\Ket{\Gamma^{6}_{i}(2)} = &\eta^1_i \Ket{\begin{array}{c} 0 \\ \Gamma^1_i \end{array}}-\frac{\eta^{4}_{i}}{\sqrt{3}}\Ket{\begin{array}{c} \Gamma^4_i(1)-\mi \Gamma^4_i(2)  \\ -\Gamma^4_i(3)\end{array}} , 
\end{align}
\begin{align}
\Ket{\Gamma^{7}_{i}(1)} = &\eta^2_i  \Ket{\begin{array}{c}\Gamma^2_i \\ 0 \end{array}}-\frac{\eta^{5}_{i}}{\sqrt{3}}\Ket{\begin{array}{c} \Gamma^5_i(3) \\ \Gamma^5_i(1)+\mi \Gamma^5_i(2) \end{array}},  \\
\Ket{\Gamma^{7}_{i}(2)} = &\eta^2_i \Ket{\begin{array}{c} 0 \\ \Gamma^2_i \end{array}}-\frac{\eta^{5}_{i}}{\sqrt{3}}\Ket{\begin{array}{c} \Gamma^5_i(1)-\mi \Gamma^5_i(2)  \\ -\Gamma^5_i(3)\end{array}} , 
\end{align}
\begin{align}
\Ket{\Gamma^{8}_{i}(1)} = &-\eta^3_i \Ket{\begin{array}{c}0 \\ \Gamma^3_i(1) \end{array}} +\frac{\eta^4_i}{\sqrt{6}} \Ket{\begin{array}{c}\Gamma^4_i(1)-\mi \Gamma^4_i(2) \\ 2\Gamma^4_i(3) \end{array}}-\frac{\eta^5_i}{\sqrt{2}} \Ket{\begin{array}{c}\Gamma^5_i(1)+\mi \Gamma^5_i(2) \\ 0 \end{array}},  \\
\Ket{\Gamma^{8}_{i}(2)} = &-\eta^3_i \Ket{\begin{array}{c}  \Gamma^3_i(2) \\0 \end{array}} -\frac{\eta^4_i}{\sqrt{2}} \Ket{\begin{array}{c}0 \\\Gamma^4_i(1)-\mi \Gamma^4_i(2) \end{array}} - \frac{\eta^5_i}{\sqrt{2}} \Ket{\begin{array}{c} -2\Gamma^5_i(3) \\ \Gamma^5_i(1)+\mi \Gamma^5_i(2) \end{array}},  \\
\Ket{\Gamma^{8}_{i}(3)} = &\eta^3_i \Ket{\begin{array}{c}  0 \\ \Gamma^3_i(2) \end{array}} +\frac{\eta^4_i}{\sqrt{2}} \Ket{\begin{array}{c} \Gamma^4_i(1)+\mi \Gamma^4_i(2) \\ 0 \end{array}} + \frac{\eta^5_i}{\sqrt{6}} \Ket{\begin{array}{c} \Gamma^5_i(1)-\mi \Gamma^5_i(2)  \\ 2\Gamma^5_i(3)\end{array}},  \\
\Ket{\Gamma^{8}_{i}(4)} = &\eta^3_i \Ket{\begin{array}{c}\Gamma^3_i(1) \\ 0  \end{array}} -\frac{\eta^4_i}{\sqrt{6}} \Ket{\begin{array}{c}-2\Gamma^4_i(3) \\ \Gamma^4_i(1)+\mi \Gamma^4_i(2) \end{array}}+\frac{\eta^5_i}{\sqrt{2}} \Ket{\begin{array}{c} 0 \\ \Gamma^5_i(1)-\mi \Gamma^5_i(2) \end{array}}.
\end{align}

In the states above we used the notation $ \Ket{\begin{array}{c}\Gamma^\alpha_i \\ \Gamma^\beta_i \end{array}} = \Ket{\Gamma^\alpha_i}\Ket{\uparrow} +  \ket{\Gamma^\beta_i}\Ket{\downarrow} $.

\section{Definition of the matrices X}
\label{App_Def_X}

For the derivation of the matrices $\mathrm{X}$ it is convenient to define the operator 
$\hat{\mathcal{P}}^{\Gamma^\mu}_{\gamma\gamma'}$ as follows \cite{Dresselhaus2008}:
\begin{equation}
\hat{\mathcal{P}}^{\Gamma^\mu}_{\gamma\gamma'}=\frac{l_\mu}{h}\sum\limits_{g} \left( \mathcal{D}^{\Gamma^\mu}_{\gamma\gamma'} (g) \right)^\ast \hat{P}_g,
\end{equation}
where $l_\mu$ denotes the dimension of $\mathcal{D}^{\Gamma^{\mu}}_{\gamma\gamma'}$, $h$ the order of the group, and the asterisk complex conjugation.
This operator has the property to project any quantity $\hat{\mathcal{O}}^\kappa_\lambda$ onto $\hat{\mathcal{O}}^\mu_\gamma$ provided 
$\mu = \kappa$ and $\gamma' = \lambda$, more precisely: 
$\hat{\mathcal{P}}^{\Gamma^\mu}_{\gamma\gamma'}\hat{\mathcal{O}}^\kappa_\lambda=\hat{\mathcal{O}}^\mu_\gamma \,\delta_{\mu\kappa}\delta_{\gamma'\lambda}$.
Applying $\hat{\mathcal{P}}^{\Gamma^\mu}_{\gamma\gamma}$ to the Hamiltonian (\ref{eq_hamiltonian}) twice, we obtain:
\begin{align}
\hat{\mathcal{P}}^{\Gamma^\mu}_{\gamma\gamma} \left( \hat{\mathcal{P}}^{\Gamma^\mu}_{\gamma\gamma}\hat{\mathcal{H}} \right) & = 
\hat{\mathcal{P}}^{\Gamma^\mu}_{\gamma\gamma}\sum\limits_{o}\mathcal{K}^{\mu}_{o; \gamma} \hat{\mathcal{O}}^{\mu}_{o; \gamma} \nonumber \\
 & = \sum\limits_{\substack{o, ij \\\alpha \beta\\\delta\delta'}}\mathcal{K}^{\mu}_{o; \gamma} \mathcal{O}^{\alpha\beta;\mu}_{o; \gamma;ij;\delta\delta'}
 \left( \hat{\mathcal{P}}^{\Gamma^\mu}_{\gamma\gamma} \Ket{\Gamma^\alpha_{i}(\delta)}\bra{\Gamma^\beta_j (\delta')} \right) .
\end{align}
According to Eq.~(\ref{eq_spatial_basis}) each symmetry operation applied to a basis state $\Ket{\Gamma^\alpha_{i}(\delta)}$ leaves the 
band index $i$ as well as the index $\mu$ of the irreducible representation unchanged. If the irreducible representation $\Gamma^\mu$ is 
contained in the direct product of $\Gamma^\alpha \otimes \Gamma^\beta$, the above expression can be written as follows:
\begin{align}
\hat{\mathcal{P}}^{\Gamma^\mu}_{\gamma\gamma}\cH^{\alpha\beta}_{ij} &=\frac{l_\mu}{h}\sum\limits_{\substack{o, g\\ \delta\delta'\\\lambda \lambda'}}
\left( \mathcal{D}^{\Gamma^{\mu}}_{\gamma\gamma} (g) \right)^\ast \mathcal{D}^{\Gamma^{\alpha}}_{\lambda\delta}(g)
\left( \mathcal{D}^{\Gamma^{\beta}}_{\lambda'\delta'} (g) \right)^\ast \mathcal{K}^{\mu}_{o; \gamma} \mathcal{O}^{\alpha\beta;\mu}_{o; \gamma;ij;\delta\delta'}\Ket{\Gamma^\alpha_{i}(\lambda)}\bra{\Gamma^\beta_j (\lambda')} \nonumber \\ 
& \equiv \sum\limits_{\substack{o\lambda \lambda'}} \mathcal{K}^{\mu}_{o; \gamma}  \mathcal{I}^{\alpha\beta; \mu}_{o;ij} \mathrm{X}^{\alpha\beta;\mu}_{\gamma;\lambda\lambda'}\Ket{\Gamma^\alpha_{i}(\lambda)}\bra{\Gamma^\beta_j (\lambda')},\label{eq_def_X}
\end{align}
for all combinations $ij$ and $\alpha\beta$. On the other hand, if the irreducible representation $\Gamma^\mu$ is {\em not}
contained in the direct product of $\Gamma^\alpha \otimes \Gamma^\beta$, all matrix elements of $\cH^{\alpha\beta}_{ij}$ are zero. 
Hence, each matrix $ \mathrm{X}^{\alpha\beta;\mu }_{\gamma;\lambda\lambda'}$ can be obtained from Eq.~(\ref{eq_def_X}). Note that if $\Gamma^\mu$ 
is contained in $\Gamma^\alpha \otimes \Gamma^\beta$ more than once, each appearance of $\Gamma^\mu$ corresponds to an invariant and a matrix. 
If needed, we take these mutual independent invariants and matrices formally into account by replacing $\mu \rightarrow \mu (\rho)$. 
All matrices $\mathrm{X}$ are in accordance with their definition, Eq.~(\ref{eq_def_X}), orthogonal and normalized in the following sense:
\begin{equation}
\sum\limits_{\delta\delta'}\mathrm{X}^{\alpha\beta;\mu(\rho)}_{\gamma;\delta\delta'}\mathrm{X}^{\beta\alpha;\mu'(\rho')}_{\gamma';\delta'\delta}
=\sum\limits_{\delta\delta'}\mathrm{X}^{\alpha\beta;\mu(\rho)}_{\gamma;\delta\delta'}\left(\mathrm{X}^{\alpha\beta;\mu'(\rho')}_{\gamma';\delta\delta'}\right)^\ast 
= \delta_{\mu \mu'}\delta_{\gamma \gamma'}\delta_{\rho \rho'}.
\label{eq_X_OrthoNorm}
\end{equation}
Note that the definition of the matrices X depends on the choice of the matrices $\mathcal{D}^{\Gamma^{\mu}}_{\gamma\gamma'} (g)$ which are not 
unique and can differ by a similarity transformation. This ambiguity, however, is, except for a trivial phase factor $\mathrm{e}^{\mi \phi}$, 
lifted by the choice of basis with respect to the crystallographic orientation.

All matrices X for the double group $T_d\otimes \mathcal{D}^{1/2}$, used in this work, were chosen to coincide -- up to a normalizing 
prefactor -- with the matrices of Refs.~\onlinecite{Trebin1979, Winkler2003}, except for the matrices $J_\gamma$ and $J_\gamma^3$ which 
transform according to $\Gamma^4$. In order to ensure orthogonality, (\ref{eq_X_OrthoNorm}), we defined 
$\mathrm{X}^{88; 4(a)}_{\gamma} = J_\gamma / \sqrt{5}$, and $\mathrm{X}^{88; 4(b)}_{\gamma} = [(5/3) J^3_\gamma - (41/12) J_\gamma]/\sqrt{5}$.
In addition, with respect to quantities which transform according to $\Gamma^5$, we introduced 
$\mathrm{X}^{88; 5(a)}_{1} = \{ J_y , J_z \} / \sqrt{3}$, and $\mathrm{X}^{88; 5(b)}_{1} = \{ (J_y^2 - J_z^2), J_x \} / \sqrt{3}$, 
plus cyclic permutations, respectively, where $\{A,B \}$ denotes half the anticommutator, $(AB + BA)/2$. See also appendix A,
especially table A.9, in Ref.~\onlinecite{Wanner2016}. The corresponding invariants are
denoted as $\mathcal{I}^{88; 5(a)}_{\bk; ll'}$ and $\mathcal{I}^{88; 5(b)}_{\bk; ll'}$, cf.\ Eq.~(\ref{EffHam_Dress_eq_lambda}).

\section{Irreducible tensor operators}

In the following we present the operators used in sections \ref{sec_eff_Ham_Td}--\ref{sec_s-cond_band} for the calculation of effective Hamiltonians: 
\begin{align}
\hat{\mathcal{O}}^{1}_{\bepsilon}\equiv &\hat{\mathcal{O}}^{1}_{\mathrm{V}} + \hat{\mathcal{O}}^{1}_{\mathrm{VI}}  = \frac{1}{3}(\hat{D}_{xx}+\hat{D}_{yy}+\hat{D}_{zz})   \nonumber \\
	&+  \frac{\hbar}{12m_0^2 c^2}\nabla(V_{xx}+V_{yy}+V_{zz}-2V_0 )\times \bp\cdot \bsigma,
\end{align}
\begin{align}
\hat{\mathcal{O}}^{3}_{\bepsilon}\equiv &\hat{\mathcal{O}}^{3}_{\mathrm{V}} + \hat{\mathcal{O}}^{3}_{\mathrm{VI}}  = \left( \begin{array}{c}
\frac{2D_{zz}-D_{xx}-D_{yy}}{\sqrt{6}}  \\
\frac{D_{xx}-D_{yy}}{\sqrt{2}} \\
\end{array} \right) \nonumber \\
&+\frac{\hbar}{4 m_0^2 c^2 } \left( \begin{array}{c}
\bp \times \bsigma \cdot \nabla \frac{2V_{zz}-V_{xx}-V_{yy}}{\sqrt{6}} + \frac{1}{\sqrt{6}}\nabla V_{0} \times \bp \cdot (-\sigma^x , -\sigma^y, 2\sigma^z)\\
\bp \times \bsigma \cdot \nabla \frac{V_{xx}-V_{yy}}{\sqrt{2}} +  \frac{1}{\sqrt{2}}\nabla V_{0} \times \bp \cdot (\sigma^x , -\sigma^y, 0)
\end{array} \right),
\end{align}
\begin{align}
\hat{\mathcal{O}}^{5}_{\bepsilon;1}\equiv &\hat{\mathcal{O}}^{5}_{\mathrm{V};1} + \hat{\mathcal{O}}^{5}_{\mathrm{VI};1}  = 2 D_{yz} + \frac{\hbar}{2 m_0^2 c^2 }\nabla\times\bp\cdot \bsigma 
V_{yz}  \nonumber \\
&+ \frac{\hbar}{4 m_0^2 c^2 } 
(\partial_{x}V_{0}p_{y}-\partial_{y}V_{0}p_{x})\sigma^{y}+(\partial_{z}V_{0}p_{x}-\partial_{x}V_{0}p_{z})\sigma^{z},
\end{align}
\begin{align}
\mathcal{O}^{1}_{\bk\bepsilon} = &\frac{\hbar}{2 \sqrt{3}m_0^2 c^2 }\nabla\times\left(\begin{array}{c}
V_{yz}\\
V_{zx}\\
V_{xy}
\end{array}   \right) \cdot \bsigma,  \nonumber \\
\mathcal{O}^{4}_{\bk\bepsilon 1;1} =&\frac{\hbar}{4 \sqrt{2}m_0^2 c^2 }\left( 
\partial_{z}(V_{0} + V_{yy} - V_{zz} )\sigma^y + \partial_{y}(V_{0} - V_{yy} + V_{zz} )\sigma^z  \right),\nonumber \\
\mathcal{O}^{4}_{\bk\bepsilon 2;1} =&\frac{\hbar}{4 \sqrt{2}m_0^2 c^2 }\left[ 2 \left( \partial_{x}V_{xy}\sigma^{z}-\partial_{z}V_{xy}\sigma^{x}+\partial_{x}V_{zx}\sigma^{y}-\partial_{y}V_{zx}\sigma^{x} \right)\right.\nonumber \\
 &-\left. \left( \partial_{z}V_{0}\sigma^y + \partial_{y}V_{0}\sigma^z   \right)\right].
\end{align}
For the operators with $\Gamma^4$ and $\Gamma^5$ symmetry only the first component is given, the remaining components are obtained 
by cyclic permutation of all indices.

\bibliography{Biblio}

\begin{thebibliography}{33}%
\makeatletter
\providecommand \@ifxundefined [1]{%
 \@ifx{#1\undefined}
}%
\providecommand \@ifnum [1]{%
 \ifnum #1\expandafter \@firstoftwo
 \else \expandafter \@secondoftwo
 \fi
}%
\providecommand \@ifx [1]{%
 \ifx #1\expandafter \@firstoftwo
 \else \expandafter \@secondoftwo
 \fi
}%
\providecommand \natexlab [1]{#1}%
\providecommand \enquote  [1]{``#1''}%
\providecommand \bibnamefont  [1]{#1}%
\providecommand \bibfnamefont [1]{#1}%
\providecommand \citenamefont [1]{#1}%
\providecommand \href@noop [0]{\@secondoftwo}%
\providecommand \href [0]{\begingroup \@sanitize@url \@href}%
\providecommand \@href[1]{\@@startlink{#1}\@@href}%
\providecommand \@@href[1]{\endgroup#1\@@endlink}%
\providecommand \@sanitize@url [0]{\catcode `\\12\catcode `\$12\catcode
  `\&12\catcode `\#12\catcode `\^12\catcode `\_12\catcode `\%12\relax}%
\providecommand \@@startlink[1]{}%
\providecommand \@@endlink[0]{}%
\providecommand \url  [0]{\begingroup\@sanitize@url \@url }%
\providecommand \@url [1]{\endgroup\@href {#1}{\urlprefix }}%
\providecommand \urlprefix  [0]{URL }%
\providecommand \Eprint [0]{\href }%
\providecommand \doibase [0]{http://dx.doi.org/}%
\providecommand \selectlanguage [0]{\@gobble}%
\providecommand \bibinfo  [0]{\@secondoftwo}%
\providecommand \bibfield  [0]{\@secondoftwo}%
\providecommand \translation [1]{[#1]}%
\providecommand \BibitemOpen [0]{}%
\providecommand \bibitemStop [0]{}%
\providecommand \bibitemNoStop [0]{.\EOS\space}%
\providecommand \EOS [0]{\spacefactor3000\relax}%
\providecommand \BibitemShut  [1]{\csname bibitem#1\endcsname}%
\let\auto@bib@innerbib\@empty
\bibitem [{\citenamefont {Luttinger}\ and\ \citenamefont
  {Kohn}(1955)}]{Luttinger1955}%
  \BibitemOpen
  \bibfield  {author} {\bibinfo {author} {\bibfnamefont {J.}~\bibnamefont
  {Luttinger}}\ and\ \bibinfo {author} {\bibfnamefont {W.}~\bibnamefont
  {Kohn}},\ }\href@noop {} {\bibfield  {journal} {\bibinfo  {journal} {Phys.
  Rev.}\ }\textbf {\bibinfo {volume} {97}},\ \bibinfo {pages} {869} (\bibinfo
  {year} {1955})}\BibitemShut {NoStop}%
\bibitem [{\citenamefont {Cardona}\ \emph {et~al.}(1988)\citenamefont
  {Cardona}, \citenamefont {Christensen},\ and\ \citenamefont
  {Fasol}}]{Cardona1988}%
  \BibitemOpen
  \bibfield  {author} {\bibinfo {author} {\bibfnamefont {M.}~\bibnamefont
  {Cardona}}, \bibinfo {author} {\bibfnamefont {N.}~\bibnamefont
  {Christensen}}, \ and\ \bibinfo {author} {\bibfnamefont {G.}~\bibnamefont
  {Fasol}},\ }\href@noop {} {\bibfield  {journal} {\bibinfo  {journal} {Phys.
  Rev. B}\ }\textbf {\bibinfo {volume} {38}},\ \bibinfo {pages} {1806}
  (\bibinfo {year} {1988})}\BibitemShut {NoStop}%
\bibitem [{\citenamefont {Novik}\ \emph {et~al.}(2005)\citenamefont {Novik},
  \citenamefont {Pfeuffer-Jeschke}, \citenamefont {Jungwirth}, \citenamefont
  {Latussek}, \citenamefont {Becker}, \citenamefont {Landwehr}, \citenamefont
  {Buhmann},\ and\ \citenamefont {Molenkamp}}]{Novik2005}%
  \BibitemOpen
  \bibfield  {author} {\bibinfo {author} {\bibfnamefont {E.~G.}\ \bibnamefont
  {Novik}}, \bibinfo {author} {\bibfnamefont {A.}~\bibnamefont
  {Pfeuffer-Jeschke}}, \bibinfo {author} {\bibfnamefont {T.}~\bibnamefont
  {Jungwirth}}, \bibinfo {author} {\bibfnamefont {V.}~\bibnamefont {Latussek}},
  \bibinfo {author} {\bibfnamefont {C.~R.}\ \bibnamefont {Becker}}, \bibinfo
  {author} {\bibfnamefont {G.}~\bibnamefont {Landwehr}}, \bibinfo {author}
  {\bibfnamefont {H.}~\bibnamefont {Buhmann}}, \ and\ \bibinfo {author}
  {\bibfnamefont {L.~W.}\ \bibnamefont {Molenkamp}},\ }\href@noop {} {\bibfield
   {journal} {\bibinfo  {journal} {Phys. Rev. B}\ }\textbf {\bibinfo {volume}
  {72}},\ \bibinfo {pages} {035321} (\bibinfo {year} {2005})}\BibitemShut
  {NoStop}%
\bibitem [{\citenamefont {Kalinin}\ \emph {et~al.}(2013)\citenamefont
  {Kalinin}, \citenamefont {Krishtopenko}, \citenamefont {Maremyanin},
  \citenamefont {Spirin}, \citenamefont {Gavrilenko}, \citenamefont {Biryukov},
  \citenamefont {Baidus},\ and\ \citenamefont {Zvonkov}}]{Kalinin2013}%
  \BibitemOpen
  \bibfield  {author} {\bibinfo {author} {\bibfnamefont {K.~P.}\ \bibnamefont
  {Kalinin}}, \bibinfo {author} {\bibfnamefont {S.~S.}\ \bibnamefont
  {Krishtopenko}}, \bibinfo {author} {\bibfnamefont {K.~V.}\ \bibnamefont
  {Maremyanin}}, \bibinfo {author} {\bibfnamefont {K.~E.}\ \bibnamefont
  {Spirin}}, \bibinfo {author} {\bibfnamefont {V.~I.}\ \bibnamefont
  {Gavrilenko}}, \bibinfo {author} {\bibfnamefont {A.~A.}\ \bibnamefont
  {Biryukov}}, \bibinfo {author} {\bibfnamefont {N.~V.}\ \bibnamefont
  {Baidus}}, \ and\ \bibinfo {author} {\bibfnamefont {B.~N.}\ \bibnamefont
  {Zvonkov}},\ }\href@noop {} {\bibfield  {journal} {\bibinfo  {journal}
  {Semiconductors}\ }\textbf {\bibinfo {volume} {47}},\ \bibinfo {pages} {1485}
  (\bibinfo {year} {2013})}\BibitemShut {NoStop}%
\bibitem [{\citenamefont {Wang}\ and\ \citenamefont {Zunger}(1996)}]{Wang1996}%
  \BibitemOpen
  \bibfield  {author} {\bibinfo {author} {\bibfnamefont {L.-W.}\ \bibnamefont
  {Wang}}\ and\ \bibinfo {author} {\bibfnamefont {A.}~\bibnamefont {Zunger}},\
  }\href {\doibase 10.1103/PhysRevB.54.11417} {\bibfield  {journal} {\bibinfo
  {journal} {Phys. Rev. B}\ }\textbf {\bibinfo {volume} {54}},\ \bibinfo
  {pages} {11417} (\bibinfo {year} {1996})}\BibitemShut {NoStop}%
\bibitem [{\citenamefont {\v{C}ukari\'{c}}\ \emph {et~al.}(2016)\citenamefont
  {\v{C}ukari\'{c}}, \citenamefont {Partoens}, \citenamefont {Tadi\'{c}},
  \citenamefont {Arsoski},\ and\ \citenamefont {Peeters}}]{Nemanja2016}%
  \BibitemOpen
  \bibfield  {author} {\bibinfo {author} {\bibfnamefont {N.~A.}\ \bibnamefont
  {\v{C}ukari\'{c}}}, \bibinfo {author} {\bibfnamefont {B.}~\bibnamefont
  {Partoens}}, \bibinfo {author} {\bibfnamefont {M.~v.}\ \bibnamefont
  {Tadi\'{c}}}, \bibinfo {author} {\bibfnamefont {V.~V.}\ \bibnamefont
  {Arsoski}}, \ and\ \bibinfo {author} {\bibfnamefont {F.~M.}\ \bibnamefont
  {Peeters}},\ }\href@noop {} {\bibfield  {journal} {\bibinfo  {journal} {J.
  Phys.: Condens. Matter}\ }\textbf {\bibinfo {volume} {28}},\ \bibinfo {pages}
  {195303} (\bibinfo {year} {2016})}\BibitemShut {NoStop}%
\bibitem [{\citenamefont {Kane}(1957)}]{Kane1957}%
  \BibitemOpen
  \bibfield  {author} {\bibinfo {author} {\bibfnamefont {E.~O.}\ \bibnamefont
  {Kane}},\ }\href@noop {} {\bibfield  {journal} {\bibinfo  {journal} {J. Phys.
  Chem. Solids}\ }\textbf {\bibinfo {volume} {1}},\ \bibinfo {pages} {249}
  (\bibinfo {year} {1957})}\BibitemShut {NoStop}%
\bibitem [{\citenamefont {L{\"o}wdin}(1951)}]{Loewdin1951}%
  \BibitemOpen
  \bibfield  {author} {\bibinfo {author} {\bibfnamefont {P.-O.}\ \bibnamefont
  {L{\"o}wdin}},\ }\href@noop {} {\bibfield  {journal} {\bibinfo  {journal} {J.
  Chem. Phys.}\ }\textbf {\bibinfo {volume} {19}},\ \bibinfo {pages} {1396}
  (\bibinfo {year} {1951})}\BibitemShut {NoStop}%
\bibitem [{\citenamefont {Winkler}(2003)}]{Winkler2003}%
  \BibitemOpen
  \bibfield  {author} {\bibinfo {author} {\bibfnamefont {R.}~\bibnamefont
  {Winkler}},\ }\href@noop {} {\emph {\bibinfo {title} {{Spin-Orbit Coupling
  Effects in Two-Dimensional Electron and Hole Systems}}}}\ (\bibinfo
  {publisher} {Springer},\ \bibinfo {address} {Berlin},\ \bibinfo {year}
  {2003})\BibitemShut {NoStop}%
\bibitem [{\citenamefont {Cardona}\ and\ \citenamefont
  {Pollak}(1966)}]{Cardona1966}%
  \BibitemOpen
  \bibfield  {author} {\bibinfo {author} {\bibfnamefont {M.}~\bibnamefont
  {Cardona}}\ and\ \bibinfo {author} {\bibfnamefont {F.~H.}\ \bibnamefont
  {Pollak}},\ }\href {\doibase 10.1103/PhysRev.142.530} {\bibfield  {journal}
  {\bibinfo  {journal} {Phys. Rev.}\ }\textbf {\bibinfo {volume} {142}},\
  \bibinfo {pages} {530} (\bibinfo {year} {1966})}\BibitemShut {NoStop}%
\bibitem [{\citenamefont {Richard}\ \emph {et~al.}(2004)\citenamefont
  {Richard}, \citenamefont {Aniel},\ and\ \citenamefont
  {Fishman}}]{Richard2004}%
  \BibitemOpen
  \bibfield  {author} {\bibinfo {author} {\bibfnamefont {S.}~\bibnamefont
  {Richard}}, \bibinfo {author} {\bibfnamefont {F.}~\bibnamefont {Aniel}}, \
  and\ \bibinfo {author} {\bibfnamefont {G.}~\bibnamefont {Fishman}},\
  }\href@noop {} {\bibfield  {journal} {\bibinfo  {journal} {Phys. Rev. B}\
  }\textbf {\bibinfo {volume} {70}},\ \bibinfo {pages} {235204} (\bibinfo
  {year} {2004})}\BibitemShut {NoStop}%
\bibitem [{\citenamefont {Bir}\ and\ \citenamefont {Pikus}(1974)}]{Bir1974}%
  \BibitemOpen
  \bibfield  {author} {\bibinfo {author} {\bibfnamefont {G.}~\bibnamefont
  {Bir}}\ and\ \bibinfo {author} {\bibfnamefont {G.~E.}\ \bibnamefont
  {Pikus}},\ }\href@noop {} {\emph {\bibinfo {title} {{Symmetry and
  Strain-Induced Effects in Semiconductors}}}}\ (\bibinfo  {publisher}
  {Wiley},\ \bibinfo {address} {New York},\ \bibinfo {year} {1974})\BibitemShut
  {NoStop}%
\bibitem [{\citenamefont {Bahder}(1990)}]{Bahder1990}%
  \BibitemOpen
  \bibfield  {author} {\bibinfo {author} {\bibfnamefont {T.~B.}\ \bibnamefont
  {Bahder}},\ }\href@noop {} {\bibfield  {journal} {\bibinfo  {journal} {Phys.
  Rev. B}\ }\textbf {\bibinfo {volume} {41}},\ \bibinfo {pages} {11992}
  (\bibinfo {year} {1990})},\ \bibinfo {note} {erratum: ibid. \textbf{46}, 9913
  (1992)}\BibitemShut {NoStop}%
\bibitem [{\citenamefont {Michelini}\ \emph {et~al.}(2009)\citenamefont
  {Michelini}, \citenamefont {Cavassilas}, \citenamefont {Hayn},\ and\
  \citenamefont {Szczap}}]{Michelini2009}%
  \BibitemOpen
  \bibfield  {author} {\bibinfo {author} {\bibfnamefont {F.}~\bibnamefont
  {Michelini}}, \bibinfo {author} {\bibfnamefont {N.}~\bibnamefont
  {Cavassilas}}, \bibinfo {author} {\bibfnamefont {R.}~\bibnamefont {Hayn}}, \
  and\ \bibinfo {author} {\bibfnamefont {M.}~\bibnamefont {Szczap}},\
  }\href@noop {} {\bibfield  {journal} {\bibinfo  {journal} {Phys. Rev. B}\
  }\textbf {\bibinfo {volume} {80}},\ \bibinfo {pages} {245210} (\bibinfo
  {year} {2009})}\BibitemShut {NoStop}%
\bibitem [{\citenamefont {L{\"o}wdin}(1968)}]{Loewdin1968}%
  \BibitemOpen
  \bibfield  {author} {\bibinfo {author} {\bibfnamefont {P.-O.}\ \bibnamefont
  {L{\"o}wdin}},\ }\href@noop {} {\bibfield  {journal} {\bibinfo  {journal}
  {Int. J. Quant. Chem.}\ }\textbf {\bibinfo {volume} {2}},\ \bibinfo {pages}
  {867} (\bibinfo {year} {1968})}\BibitemShut {NoStop}%
\bibitem [{\citenamefont {Brandow}(1967)}]{Brandow1967}%
  \BibitemOpen
  \bibfield  {author} {\bibinfo {author} {\bibfnamefont {B.~H.}\ \bibnamefont
  {Brandow}},\ }\href@noop {} {\bibfield  {journal} {\bibinfo  {journal} {Rev.
  Mod. Phys.}\ }\textbf {\bibinfo {volume} {39}},\ \bibinfo {pages} {771}
  (\bibinfo {year} {1967})}\BibitemShut {NoStop}%
\bibitem [{\citenamefont {Lindgren}(1974)}]{Lindgren1974}%
  \BibitemOpen
  \bibfield  {author} {\bibinfo {author} {\bibfnamefont {I.}~\bibnamefont
  {Lindgren}},\ }\href@noop {} {\bibfield  {journal} {\bibinfo  {journal} {J.
  Phys. B: At. Mol. Phys.}\ }\textbf {\bibinfo {volume} {7}},\ \bibinfo {pages}
  {2441} (\bibinfo {year} {1974})}\BibitemShut {NoStop}%
\bibitem [{\citenamefont {Fr\'esard}\ \emph {et~al.}(2012)\citenamefont
  {Fr\'esard}, \citenamefont {Hackenberger},\ and\ \citenamefont
  {Kopp}}]{Fresard2012}%
  \BibitemOpen
  \bibfield  {author} {\bibinfo {author} {\bibfnamefont {R.}~\bibnamefont
  {Fr\'esard}}, \bibinfo {author} {\bibfnamefont {C.}~\bibnamefont
  {Hackenberger}}, \ and\ \bibinfo {author} {\bibfnamefont {T.}~\bibnamefont
  {Kopp}},\ }\href@noop {} {\bibfield  {journal} {\bibinfo  {journal} {Ann.
  Phys. (Berlin)}\ }\textbf {\bibinfo {volume} {524}},\ \bibinfo {pages} {411}
  (\bibinfo {year} {2012})}\BibitemShut {NoStop}%
\bibitem [{\citenamefont {Ludwig}\ and\ \citenamefont
  {Falter}(1996)}]{Ludwig1996}%
  \BibitemOpen
  \bibfield  {author} {\bibinfo {author} {\bibfnamefont {W.}~\bibnamefont
  {Ludwig}}\ and\ \bibinfo {author} {\bibfnamefont {C.}~\bibnamefont
  {Falter}},\ }\href@noop {} {\emph {\bibinfo {title} {Symmetries in Physics:
  Group Theory Applied to Physical Problems}}}\ (\bibinfo  {publisher}
  {Springer},\ \bibinfo {address} {Berlin},\ \bibinfo {year}
  {1996})\BibitemShut {NoStop}%
\bibitem [{\citenamefont {Bradley}\ and\ \citenamefont
  {Cracknell}(2010)}]{Bradley2010}%
  \BibitemOpen
  \bibfield  {author} {\bibinfo {author} {\bibfnamefont {C.~J.}\ \bibnamefont
  {Bradley}}\ and\ \bibinfo {author} {\bibfnamefont {A.~P.}\ \bibnamefont
  {Cracknell}},\ }\href@noop {} {\emph {\bibinfo {title} {The Mathematical
  Theory of Symmetry in Solids: Representation Theory for Point Groups and
  Space Groups}}}\ (\bibinfo  {publisher} {Clarendon},\ \bibinfo {address}
  {Oxford},\ \bibinfo {year} {2010})\BibitemShut {NoStop}%
\bibitem [{\citenamefont {Dresselhaus}(2008)}]{Dresselhaus2008}%
  \BibitemOpen
  \bibfield  {author} {\bibinfo {author} {\bibfnamefont {M.~S.}\ \bibnamefont
  {Dresselhaus}},\ }\href@noop {} {\emph {\bibinfo {title} {Group Theory
  Application to the Physics of Condensed Matter}}}\ (\bibinfo  {publisher}
  {Springer},\ \bibinfo {address} {Berlin},\ \bibinfo {year}
  {2008})\BibitemShut {NoStop}%
\bibitem [{\citenamefont {Altmann}\ and\ \citenamefont
  {Herzig}(1994)}]{Altmann1994}%
  \BibitemOpen
  \bibfield  {author} {\bibinfo {author} {\bibfnamefont {S.~L.}\ \bibnamefont
  {Altmann}}\ and\ \bibinfo {author} {\bibfnamefont {P.}~\bibnamefont
  {Herzig}},\ }\href@noop {} {\emph {\bibinfo {title} {Point-Group Theory
  Tables}}}\ (\bibinfo  {publisher} {Clarendon},\ \bibinfo {address} {Oxford},\
  \bibinfo {year} {1994})\BibitemShut {NoStop}%
\bibitem [{Note1()}]{Note1}%
  \BibitemOpen
  \bibinfo {note} {The matrix representation $\protect \mathcal {D}^{\Gamma
  ^{\mu }}_{\gamma '\gamma } (g)$ can be found in textbooks such as
  Ref.~\protect \rev@citealp {Bradley2010}.}\BibitemShut {Stop}%
\bibitem [{\citenamefont {Wanner}(2016)}]{Wanner2016}%
  \BibitemOpen
  \bibfield  {author} {\bibinfo {author} {\bibfnamefont {J.}~\bibnamefont
  {Wanner}},\ }\emph {\bibinfo {title} {Acoustically Induced Spin Relaxation in
  Two-Dimensional Electronic Systems}},\ \href@noop {} {Ph.D. thesis},\
  \bibinfo  {school} {University of Augsburg} (\bibinfo {year} {2016}),\
  \bibinfo {note} {available online at
  \url{https://opus.bibliothek.uni-augsburg.de/opus4/frontdoor/index/index/docId/3407}}\BibitemShut
  {NoStop}%
\bibitem [{\citenamefont {Koster}(1963)}]{Koster1963}%
  \BibitemOpen
  \bibfield  {author} {\bibinfo {author} {\bibfnamefont {G.~F.}\ \bibnamefont
  {Koster}},\ }\href@noop {} {\emph {\bibinfo {title} {Properties of the
  Thirty-Two Point Groups}}}\ (\bibinfo  {publisher} {MIT Press},\ \bibinfo
  {year} {1963})\BibitemShut {NoStop}%
\bibitem [{\citenamefont {Trebin}\ \emph {et~al.}(1979)\citenamefont {Trebin},
  \citenamefont {R{\"o}ssler},\ and\ \citenamefont {Ranvaud}}]{Trebin1979}%
  \BibitemOpen
  \bibfield  {author} {\bibinfo {author} {\bibfnamefont {H.-R.}\ \bibnamefont
  {Trebin}}, \bibinfo {author} {\bibfnamefont {U.}~\bibnamefont {R{\"o}ssler}},
  \ and\ \bibinfo {author} {\bibfnamefont {R.}~\bibnamefont {Ranvaud}},\
  }\href@noop {} {\bibfield  {journal} {\bibinfo  {journal} {Phys. Rev. B}\
  }\textbf {\bibinfo {volume} {20}},\ \bibinfo {pages} {686} (\bibinfo {year}
  {1979})}\BibitemShut {NoStop}%
\bibitem [{\citenamefont {Dresselhaus}(1955)}]{Dresselhaus1955}%
  \BibitemOpen
  \bibfield  {author} {\bibinfo {author} {\bibfnamefont {G.}~\bibnamefont
  {Dresselhaus}},\ }\href@noop {} {\bibfield  {journal} {\bibinfo  {journal}
  {Phys. Rev.}\ }\textbf {\bibinfo {volume} {100}},\ \bibinfo {pages} {580}
  (\bibinfo {year} {1955})}\BibitemShut {NoStop}%
\bibitem [{\citenamefont {Suzuki}\ and\ \citenamefont
  {Hensel}(1974)}]{Suzuki1974}%
  \BibitemOpen
  \bibfield  {author} {\bibinfo {author} {\bibfnamefont {K.}~\bibnamefont
  {Suzuki}}\ and\ \bibinfo {author} {\bibfnamefont {J.}~\bibnamefont
  {Hensel}},\ }\href@noop {} {\bibfield  {journal} {\bibinfo  {journal} {Phys.
  Rev. B}\ }\textbf {\bibinfo {volume} {9}} (\bibinfo {year}
  {1974})}\BibitemShut {NoStop}%
\bibitem [{\citenamefont {Zhang}(1994)}]{Zhang1994}%
  \BibitemOpen
  \bibfield  {author} {\bibinfo {author} {\bibfnamefont {Y.}~\bibnamefont
  {Zhang}},\ }\href@noop {} {\bibfield  {journal} {\bibinfo  {journal} {Phys.
  Rev. B}\ }\textbf {\bibinfo {volume} {49}},\ \bibinfo {pages} {14352}
  (\bibinfo {year} {1994})}\BibitemShut {NoStop}%
\bibitem [{\citenamefont {Bernevig}\ and\ \citenamefont
  {Zhang}(2005)}]{Bernevig2005}%
  \BibitemOpen
  \bibfield  {author} {\bibinfo {author} {\bibfnamefont {B.~A.}\ \bibnamefont
  {Bernevig}}\ and\ \bibinfo {author} {\bibfnamefont {S.-C.}\ \bibnamefont
  {Zhang}},\ }\href@noop {} {\bibfield  {journal} {\bibinfo  {journal} {Phys.
  Rev. B}\ }\textbf {\bibinfo {volume} {72}},\ \bibinfo {pages} {115204}
  (\bibinfo {year} {2005})}\BibitemShut {NoStop}%
\bibitem [{\citenamefont {Norman}\ \emph {et~al.}(2010)\citenamefont {Norman},
  \citenamefont {Trowbridge}, \citenamefont {Stephens}, \citenamefont
  {Gossard}, \citenamefont {Awschalom},\ and\ \citenamefont
  {Sih}}]{Norman2010}%
  \BibitemOpen
  \bibfield  {author} {\bibinfo {author} {\bibfnamefont {B.~M.}\ \bibnamefont
  {Norman}}, \bibinfo {author} {\bibfnamefont {C.~J.}\ \bibnamefont
  {Trowbridge}}, \bibinfo {author} {\bibfnamefont {J.}~\bibnamefont
  {Stephens}}, \bibinfo {author} {\bibfnamefont {A.~C.}\ \bibnamefont
  {Gossard}}, \bibinfo {author} {\bibfnamefont {D.~D.}\ \bibnamefont
  {Awschalom}}, \ and\ \bibinfo {author} {\bibfnamefont {V.}~\bibnamefont
  {Sih}},\ }\href@noop {} {\bibfield  {journal} {\bibinfo  {journal} {Phys.
  Rev. B}\ }\textbf {\bibinfo {volume} {82}},\ \bibinfo {pages} {081304}
  (\bibinfo {year} {2010})}\BibitemShut {NoStop}%
\bibitem [{\citenamefont {Kato}\ \emph {et~al.}(2004)\citenamefont {Kato},
  \citenamefont {Myers}, \citenamefont {Gossard},\ and\ \citenamefont
  {Awschalom}}]{Kato2004}%
  \BibitemOpen
  \bibfield  {author} {\bibinfo {author} {\bibfnamefont {Y.}~\bibnamefont
  {Kato}}, \bibinfo {author} {\bibfnamefont {R.~C.}\ \bibnamefont {Myers}},
  \bibinfo {author} {\bibfnamefont {A.~C.}\ \bibnamefont {Gossard}}, \ and\
  \bibinfo {author} {\bibfnamefont {D.~D.}\ \bibnamefont {Awschalom}},\
  }\href@noop {} {\bibfield  {journal} {\bibinfo  {journal} {Nature}\ }\textbf
  {\bibinfo {volume} {427}},\ \bibinfo {pages} {50} (\bibinfo {year}
  {2004})}\BibitemShut {NoStop}%
\bibitem [{\citenamefont {Hru\v{s}ka}\ \emph {et~al.}(2006)\citenamefont
  {Hru\v{s}ka}, \citenamefont {Kos}, \citenamefont {Crooker}, \citenamefont
  {Saxena},\ and\ \citenamefont {Smith}}]{Hruska2006}%
  \BibitemOpen
  \bibfield  {author} {\bibinfo {author} {\bibfnamefont {M.}~\bibnamefont
  {Hru\v{s}ka}}, \bibinfo {author} {\bibfnamefont {{\v{S}}.}~\bibnamefont
  {Kos}}, \bibinfo {author} {\bibfnamefont {S.~A.}\ \bibnamefont {Crooker}},
  \bibinfo {author} {\bibfnamefont {A.}~\bibnamefont {Saxena}}, \ and\ \bibinfo
  {author} {\bibfnamefont {D.~L.}\ \bibnamefont {Smith}},\ }\href@noop {}
  {\bibfield  {journal} {\bibinfo  {journal} {Phys. Rev. B}\ }\textbf {\bibinfo
  {volume} {73}},\ \bibinfo {pages} {075306} (\bibinfo {year}
  {2006})}\BibitemShut {NoStop}%
\end{thebibliography}%

\end{document}